\documentclass[prd,aps,floats,floatfix,eqsecnum,nofootinbib,12pt]{revtex4-1}
\usepackage{verbatim,graphicx,amssymb,amsbsy,bm,amsmath,rotating,epsfig}
\usepackage{psfrag}
\usepackage{hyperref}
\usepackage{fontenc}
\usepackage[latin1]{inputenc}
\usepackage{pstricks,pst-node,pst-text,pst-3d}
\usepackage{textcomp}
\newcommand{\be}{\begin{equation}}
\newcommand{\ee}{\end{equation}}
\newcommand{\bea}{\begin{eqnarray}}
\newcommand{\eea}{\end{eqnarray}}

\begin{document}

\title{The Classical-Quantum Duality of Nature.\\
New Variables for Quantum Gravity}
\author{Norma G. SANCHEZ}
\affiliation{
LERMA CNRS UMR 8112 Observatoire de Paris PSL Research University, 
Sorbonne Universit\'e UPMC Paris VI,\\
 61, Avenue de l'Observatoire, 75014 Paris, France}
\date{\today}

\begin{abstract}
{\bf Abstract}: The classical-quantum duality at the basis of quantum theory is here extended to 
the Planck scale domain. The classical/semiclassical gravity (G) domain
is dual (in the precise sense of the classical-quantum duality) to the quantum (Q) 
elementary particle domain: $ O_Q = o_P^2 O_G^{-1} $,  $o_P$ being the Planck scale.
This duality is {\it universal}. From the gravity (G) and quantum 
(Q) variables $(O_G, O_Q)$, we define 
new (QG) quantum gravity variables $O_{QG} = (1/2) (O_G + O_Q)$  which include 
all (classical, semiclassical and quantum gravity) domains passing by the 
Planck scale and the elementary particle domain. The QG variables are more 
{\it complete} than the usual ($O_Q$, $O_G$) ones which cover only one domain
(Q or G ).{\it Two} $O_{G}$ or $O_{Q}$ values $(\pm )$ are needed for 
each value of $O_{QG}$ (reflecting the {\it two} different and {\it dual} 
ways of reaching the Planck scale). We perform the 
complete analytic extension of the QG variables through analytic (holomorphic)
mappings which preserve the ligth-cone structure. This allows us 
to reveal the {\it classical-quantum 
duality} of the Schwarzschild-Kruskal space-time: exterior regions
are classical or semiclassical while the interior is totally
{\it quantum}: its  boundaries being the Planck scale. Exterior and interior
lose their difference near the horizon: 
four Planck scale hyperbolae border the horizons as a {\it quantum dressing}
or width:{\it "l'horizon habill\'{e}"}. QG variables are naturally
invariant under  $O_G \rightarrow O_Q$.
Space-time reflections, antipodal symmetry and PT or CPT symmetry are contained 
in the QG symmetry, which also shed insight into the  global  properties
of the Kruskal manifold and its present renewed interest. Further results are 
presented in another paper.\\
Norma.Sanchez@obspm.fr\\
\url{https://chalonge-devega.fr/sanchez}
\end{abstract}
\maketitle
\tableofcontents

\section{Introduction and Results}

Nature is Quantum. That means that the real and complete laws of nature are those 
of quantum physics. Classical behaviours and domains are particular cases, 
limiting situations or approximations. Classical gravity, 
and thus successful General Relativity are incomplete (non quantum) theories 
and must be considered as a particular approximation from a more complete theory yet to achieve. 

A complete quantum theory should include and account for the physics at the Planck scale and 
domain. 
As is known, the task to complete such a theory is far from being easy 
and follows in summary at least two different paths: \textbf{(i)} Starting from gravity, 
that is General Relativity and quantizing it (by applying
the different quantization -perturbative and non perturbative- procedures,
with the by now well known shortcomings and developpements and its rich bibliography -is not our 
aim here to review it-: canonical (hamiltonian) quantization, Wheeler-de Witt formalisms,  path 
integral (euclidean) gravity, loop gravity, string theory....). 
\textbf{(ii)} Starting from Quantum theory and trying to extend it to the Planck scale domain.  
(instead of going from classical gravity to quantum gravity). 
Of course, in constructing the road (ii) many of the lessons from road (i) are most
 useful. One tractable and well posed piece of work is semiclassical gravity,
 in its several degrees of quantization: QFT in curved backgrounds and its (perturbative 
and non perturbative) back reaction (semiclassical Einstein Equations), and the path integral 
approach. Examples are the Hawking radiation, 
the early universe inflation and the primordial quantum fluctuations, 
seeds of the structure in the Universe imprinted in the 
 CMB temperature anisotropies and polarization. Moreover, 
 the expanding quantum cosmological vacuum could be the source of the present acceleration 
 of the universe (dark energy) compatible with a cosmological constant.

\medskip

Let us recall that one of the starting milestones of quantum theory is the concept 
of classical-quantum (wave-particle) duality. In ref [1] we extended this concept 
to include the quantum gravity domain, ie wave-particle-gravity
duality. We had set up the relevant scales characteristic of the classical 
gravity regime and related it to the 
semiclassical and quantum gravity regimes (being QFT or strings) refs [1], [2]. 
The quantum (Compton) wave length
\begin{equation}
L_Q =\frac{\hbar} {M c},
 \end{equation}

in the presence of gravity means 
\begin{equation}
L_Q=\frac{l_{P}^2} {L_{G}},  \qquad  (L_{G}=\frac{G}{c^2}\, M)
\end{equation}

 $L_{G}$ and $l_{P}$ being the gravitational 
length  and Planck scale length 
respectively. As the quantum behaviour is dual of the classical behaviour 
(through $\hbar$, \,$c $), the dual of the classical gravity regime is the
quantum gravity regime (through $\hbar $,\, $G$,\, $c$), 
(in string theory, this classical-quantum duality 
is through $\hbar$ ,\, $\alpha '$,\, $c$) refs [1],[2]. 

The {\it quantum mass} scale is the {\it dual mass} 
\be
M_{Q}=\frac{m_{P}^2} {M}, \qquad \mbox{$(m_{P}$ being the Planck~mass).}
\ee
Its corresponding temperature scale (energy in units of Boltzmann's constant) 
is precisely the Hawking temperature 
\be
T_{eQ}=\frac{1}{2\pi k_B} M_{Q}\, c^2,
\ee
which is a quantum temperature, mesure of the Compton length (in units of $k_B$). 
The set of quantities $O_{G}=(L_{G},\, M,\, \mathcal{K}_{G},\, 
T_e)$ characteristic of the classical gravity regime, 
(here denoting for instance size, mass, surface gravity (or gravity acceleration),  
usual temperature respectively), and the corresponding set of quantities in the quantum 
regime $O_Q=(L_Q,\, M_{Q},\, \mathcal{K}_Q,\, T_{eQ})$ are \textbf {dual} of each other, 
(in the precise sense of the wave-particle duality), ie:
\begin{equation}
O_{G}=o_{P}^2\, O_Q^{-1}
\end{equation}

This relation holding for each quantity in the set. $O_{G}$ and 
$O_Q$ are the same conceptual physical quantities in the different 
(classical/semiclassical and quantum) gravity regimes respectively . $o_{P}$ 
being the corresponding quantity at the Planck scale purely 
depending of ($\hbar ,\, c,\, G)$. $O_Q$ standing for the concepts 
of quantum size $ L_Q$, quantum mass $ M_Q$, quantum acceleration $ K_Q = c^2/L_Q$, 
and quantum temperature $T_{eQ}$ for instance. 

\medskip

This is not an assumed or conjectured duality. 
As the wave-particle duality, 
this classical/semiclassical-quantum gravity duality does not relate to the number of 
dimensions, nor to a particular or imposed symmetry of the background manifold or space-time. 
These duality relations extend too to the microscopic density of states and 
entropies ($\rho,\, S$) ref [1]. 

\medskip

[For the sake of completeness, (although we will not do string theory here) we recall 
that for strings we found similarly 
$O_{s}=o_s^2\, O_Q^{-1}$ with $\alpha '$ instead of $G/c^2$ 
and $o_s^2$ purely depending on 
($\hbar ,\, \alpha ',\, c$) . $O_s=(L_s,\, M_s,\, \mathcal{K}_s,\, T_{es})$ 
denoting here the characteristic string size, string mass, string 
surface gravity or acceleration and string temperature of the system under consideration, 
($T_s=\frac{1}{2\pi k_B}\, M_s\, c^2$), which are in general different from 
the usual flat space string expressions, (in particular they can be equal), 
refs [1], [2], [3-6]. Again, this is not an assumed or conjectured duality or a 
{\it at priori} imposed symmetry: The results of QFT and 
quantum string dynamics in curved backgrounds remarkably show these relations
refs [1], [2]-[6].] 

\bigskip

In this paper we go further in the classical-quantum duality to include 
the Planck scale domain.
Each of the sides of the duality above described, for ex. Eq (1.5) and previuosly treated 
accounts for only one domain separately: 
classical {\it or} quantum, $ O_G$ or $O_Q$ but not for both domains together. And the 
Planck scale $o_P$  
is just a fixed scale and a appropriate unit, but not a variable. In this paper 
we define appropriated quantum 
gravity (QG) variables fully taking into account all domains, classical and quantum 
including gravity and their duality properties.
From the variables $(O_G, O_Q)$, we construct QG variables $O_{QG}$ which in 
units of the corresponding Planck scale magnitude $o_P$ simply read:
\be
O_{QG} = \frac{1}{2}\;(\; O_{G} + O_{Q}\;),\quad 
\frac{O_{QG}}{o_P} \;\equiv \; O =  \frac{1}{2}\;(\; x + \frac {1}{x}\;),
\qquad x = \frac{O_G}{o_P} = \frac{o_P}{O_Q}
\ee
and which automatically endowe {\it the symmetry}: 
$$ O (1/x) = O(x), \quad \mbox{and}\quad  O(x = 1) = 1 \;\; \mbox{Planck scale.} $$
Fig $1$ shows the QG variable  $X \equiv L_{QG} / l_P$ against $x$. 
For comparison, each of its dual components ($L_Q$ and $L_G$) are included too.

\medskip

The QG variables are {\it complete}: they describe both the classical, semiclassical 
and quantum gravity domains passing by the Plank scale and the elementary particle domain as 
well. 
{\it Two} values $ x_{\pm}$ of the usual variables $O_G$ or $O_Q$
are necessary for each variable {QG}. The $(+)$ and $(-)$ branches precisely 
correspond to the two different and dual ways of reaching the Planck scale: 
from the quantum elementary particle domain ($0\leq x \leq 1$) on one side and 
from the classical/semiclassical gravity domain ($1 \leq  x \leq \infty$) 
on the other side. 
There is thus a {\it duality} between the quantum domain of elementary particles 
(without gravity) and the classical gravity domain of macroscopic masses (without the quantum 
theory). 

\medskip

In other words: The quantum domain of elementary particles 
is {\it dual} (in the precise meaning of the classical-quantum duality) 
of the classical gravity domain. This duality is {\it universal}. 
As the wave-particle duality, it does not relate to the number of dimensions 
or compactifications, nor to any particular imposed symmetry or manifold /space-time. 

\medskip

Similarly, for the QG momentum $P$ and QG time $T$  we have:
\be
\frac{T_{QG}}{t_P} \;\equiv \; T =  \frac{1}{2}\;(\; t - \frac {1}{t}\;), 
\quad t \equiv \frac{T_G}{t_P}
\ee
\be
\mbox{which satisfy} \;\; T (1/t) = -T(t), \;\; T(1) = 0 \;\; \mbox {at the Planck scale}. 
\ee 
QG light-cone type variables can be defined and
we perform the complete analytic extension of the QG variables through analytic (or holomorphic) 
mappings which preserve the light-cone type structure. Fig 2 shows the complete
QG $(X, T)$ manifold.
Four patches I, II, II, IV are necessary to cover 
the full manifold QG.
The $(X,T)$ QG variables are Kruskal's type complete coordinates while the $(x, t)$ 
variables are like
the incomplete Schwarzschild's (or Rindler) ones. 
The analogy is even more manifest in terms of star coordinates $x = \exp{(\kappa x*)}$ 
related to the Schwarzschild coordinates $(r, t*)$ by: $x = \sqrt{1- 2 \kappa r}\;\exp{(\kappa r)}, t = \exp{(\kappa t*)}$. 

\medskip

This allows us to reveal the {\it classical-quantum duality} of the 
Schwarzschild-Kruskal space-time. 
The regions which are exterior to the Planck scale hyperbolae $ X^2-T^2 = \pm 1$, 
(exterior regions I and III), are classical or semiclassical gravity domains while 
the interior regions are totally {\it quantum}, 
from the Planck scale $ X^2-T^2 = \pm 1$ hyperbolae \, passing the null horizons $X = \pm T$ \; 
(at $x=0, 2\kappa r = 1$), till the Planck scale again (at $x = 1, r = 0$): $ T^2-X^2 = \pm 1$, 
(regions II and IV). The exterior region I and the interior region II are dual 
of each other in the precise sense of the classical-quantum duality. 
This is also reflected in the duality symmetry $X (1/t) = X (t),\;T (1/t) = -T(t)$ 
of the Kruskal manifold which shows up here.

\medskip

The classical-quantum duality maps 
the internal region $ x \in (0, 1)$ into the external classical domain 
$x \in (1, \infty)$  through the Planck scale $(x=1, t=1)$ and conversely. 
The two identical halves or "worlds"  (I, II) and (III, IV) 
which are space-time reflections and  antipodally symmetric of each other are 
{\it classical-quantum duals} of each other. 
The QG duality symmetry includes
the classical/semiclassical antipodal space-time symmetry and supports 
{the antipodal identification} which translates into the PT or CPT invariance of the 
quantum theory refs [7],[8]. In a different approach, 
't Hooft required recently the antipodal identification for the black hole quantum theory
and its unitarity refs [8],[9], and our QG results here support such results.

\medskip

In the interior regions the Planck scale is reached at the loci $T^2 -X^2 = \pm 1$ 
($r=0$) which are bordering the null horizons $X = \pm T$ (at $2 \kappa r = 1$) 
and asymptotically reaching them for $T\rightarrow  \pm \infty $.
The {\it quantum region} extends beyond the horizons and is delimitated by 
the Planck scale hyperbolae $X^2 -T^2 = \pm 1$ and $T^2 - X^2 = 
\pm 1$. These hyperbolae are acting like a {\it quantum dressing} or quantum 
width for the horizon, let us call it {\it l'horizon habill\'{e}} (dressed horizon).
$1$ is here $l_P^2$. We see that {\it exterior and interior lose its difference
near the horizons $X =\pm T$}. 
Although, is possible to use these words in scales larger than the Planck scale.

There are {\it two dual} ways of reaching the horizon and the Planck scale: from the exterior 
(classical/semiclassical) and from the interior (quantum) regions. This appears like a 
{\it"splitting" or shifting} of the null horizons $X = \pm T$ 
into the hyperbolae $X^2 - T^2 = \pm 1$.  We find a similar result and other new
related results when promoting $(X,T)$ to quantum non-commutative coordinates as we do in 
ref [10].

\medskip

Let us recall here that quantum field and string dynamics near horizons
required quantum "shifts" or widths of the horizon 
of the order of the Planck scale, refs [11],[12],[13]. 
In QFT, the horizon shift appears as produced by the non-perturbative back reaction
and the dynamics of the ingoing and outgoing modes crossing the horizon ref [11]. 
In string theory the horizon width appears as due to the non-zero (Planck scale) 
size of the string refs [12],[13]. 
We find interesting that in a different approach, 
and solely from our QG variables, a horizon "quantum dressing" or width 
does appear (and without searching for it)   
supporting the quantum nature of this space-time region. The QG 
variables include naturally the Planck scale.
Other related results are reported in ref [10].

\medskip 
 
Of course, the idea of a continuum space-time is a classical 
notion but it could be considered as induced from a QG state within 
the Planck scale domain. The QG variables or coordinates cover all the domains, 
classical and quantum, with and without gravity. In the classical and 
semiclassical gravity domains, they can be used 
directly as space-time variables or coordinates as we do here. 
In the QG Planck scale domain and beyond they 
can be considered as expectation values in a quantum state 
or they can be used to construct quantum operators as we do in ref [10].

\medskip

By promoting the QG variables to quantum (non conmutative) operators 
further insight into the 
quantum space-time structure is obtained and the antipodal
identification of the two copies of the Kruskal global manifold 
is further supported by the quantum theory, ref [10].

\medskip

QG variables can be also considered in phase-space 
with their full analytic extension to all values or copies.  
Comparison of the QG variables with the complete Q-variables of 
the harmonic oscillator is enlighting.
    
\bigskip

This paper is organized as follows: In Section II we set up the new  QG variables 
for Quantum Gravity which fully take into account 
the classical-quantum (wave-particle) duality including gravity. 
Section III deals with the complete analytic extension of the QG variables, 
and their symmetry properties.
Section IV deals with Schwarzschild-Kruskal black hole manifold, its classical-quantum duality 
properties, symmetries and extensions in this new context. 
In Section V we present our remarks and conclusions.

\section {New Variables for Quantum Gravity (QG) and their Duality Properties}

In classical gravity, length and mass are  proportional (through $G$ and $c$), 
(gravitational length):
\begin{equation}
L_{G} = \frac{G}{c^2} M
\end {equation}
In quantum physics, length and mass are inversely proportional (through, 
$\hbar$ and $c$), (Compton length):
\begin{equation}
L_{Q} = \frac{\hbar}{c M}
\end{equation}
The equality $L_{G} $ = $ L_{Q}$ yielding the Planck scale mass $m_P$ (and length $l_P$)
which satisfies by definition
\begin{equation}
\frac{l_P} {m_P} = \frac{G}{c^2}     \quad ,    \quad   l_P\; m_P = \frac{\hbar}{c},  
\end{equation}
exhibiting the classical - quantum duality between  the gravity tension $G/c^2$, 
and the quantum action $ \hbar/c$. 

\bigskip

Each length separately $ L_G$ or $L_Q$, (and their respective associated energy scales)
accounts only for one physical domain 
(classical gravity or quantum theory) but not for both of them. 
Planck scale $l_P$ or $ m_P$ accounts for the quantum gravity domain 
but it is just a fixed scale 
and an appropiated unit, not a variable. 

\bigskip

In the known elementary particle physics domain, all masses are too much smaller 
with respect to the Planck mass $m_P$, which is a enormously heavy mass 
for a elementary particle, $L_Q$ is 
much larger than $l_P$, but  $L_G$ is much smaller than $l_P$ for the elementary particles  
although not being quantum gravity objects:  
\begin{equation}
Elementary ~ Particles,\; L_G < L_Q :\; 0 \leq M < m_P, \quad  0\leq  L_G <  l_P , 
\quad l_P < L_Q \leq \infty
\end{equation}	

On the other hand, in the classical gravitational domain, masses are much larger than $m_P$ which 
is a ridicously small mass for the macroscopic bodies, $L_G$ is larger than 
$l_P$ but their $L_Q$ is inside $l_P$,  although not being quantum gravitational objects:
\begin{equation}
Classical ~ Gravity, L_G  > L_Q : \; m_P < M < \infty, \quad  l_P < L_G < \infty, 
\quad  0 \leq L_Q < l_P 
\end{equation}

An appropiated length variable for quantum gravity $(QG)$ fully taking into account 
{\it both domains}, classical and quantum is the following:
\begin{equation}
L_{QG} =  \frac {1}{2}\; ( L_G + L_Q ), \quad  \mbox{that is:}\quad 
\frac{L_{QG}}{l_P} = \frac {1}{2} \;(\frac{M}{m_P} + \frac{m_P}{M} )
\end{equation}
 
$L_{QG}$ contains {\it both} $L_G$ and $L_Q$ and their duality through the Planck scale:

\begin{itemize}
 
\item {For \; $m_P < M \leq \infty$: 
$\quad$ $L_{QG} \simeq L_G$, $\quad$ \; $L_G > L_Q$}
\item {For  \; $0 \leq M < m_P: \quad L_{QG} \simeq L_Q, \quad \; L_Q > L_G$}
\item {For \; $M = m_P :  L_{QG} = 1 = L_Q = L_G= l_P$}

\end{itemize}
 
Or, in dimensionless variables: 
\begin{equation}
X = \frac {1}{2} \;( x + \frac{1}{x} )  , 
\qquad   X \equiv \frac{L_{QG}}{l_P} ,   
\qquad x \equiv \frac{M}{m_P} 
\end{equation} 
These are pure numbers, all magnitudes are defined with respect to their 
corresponding Planck scale expressions, and the variable $X$ 
is invariant under the transformation $ x\rightarrow x ^{-1}$:
\be X\;({x}^{-1}) = X\;(x), \quad X\;(x >> 1) = x , 
\quad   X\;(x << 1) = {x}^{-1}, \quad X\;(x = 1) = 1 \ee
$X$ endowes both the classical  $(x >> 1)$ and quantum  $(x << 1)$ 
domains and the Planck domain $(x \approx 1)$ as well. 
The physical classical-quantum duality intrinsically 
manifests in the QG variable $ X$. Similarly, 
we have for the QG mass variable $M_{QG}$:
\be
\frac{M_{QG}}{m_P} \;=\; \frac{1}{2} \;(\frac{M}{m_P} + \frac{m_P}{M} )\;
=\;  \frac {1}{2} \; (x + \frac{1}{x}) 
\ee
A full set of  QG-variables, as Energy $E_{QG}= M_{QG}\;c^2$ , 
Temperature $T_{eQG}= M_{QG}\; c^2/2\pi k_B$, ($k_B$ being the Boltzmann constant),
surface gravity (or gravity acceleration) $K_{QG} = c^2 / X_{QG}$ 
(or other relevant magnitudes) 
and their corresponding dimensionless expressions follow in analogous way:
\be
\frac{L_{QG}}{l_P} = \frac{M_{QG}}{m_P} = \frac{K_{QG}}{k_P} = 
\frac{T_{e QG}}{t_{eP}}\; =\; \frac {1}{2}(x + \frac{1}{x})  
\ee
$\kappa_P$ and $ t_{eP} $ being the Planck acceleration 
(Planck surface gravity) and the Planck temperature respectively:
\begin{equation}
\kappa _{P}\equiv  \frac{c^2}{l_{P}}, \quad \quad  {t}_{eP}\equiv \frac{1}{2\pi k_B} m_{P}\,c^2
\end{equation}

Several comments are in order here:
\begin{itemize}
\item {QG variables contain as limiting cases, the quantum (wave-particle) domain $x << 1$,
 the classical/semiclassical gravity ($x >> 1$) domain and the Planck ($x\sim 1$) domain as well.}
\item {QG-variables are more "complete" than the usual incomplete variables $x$ which 
only describe one domain (Q or G) and there are two values $ x_{\pm}$ for each  $X_{QG}$.}
\end{itemize}

The completion of the complete manifold of QG variables requires several "patches" 
or analytic extensions to cover the full sets $X \geq 1$ or $X \leq 1$:
\begin{equation}
x_{\pm} =  X \pm\sqrt{X^2 - 1}, \quad X \geq 1; \qquad
x_{\pm } =  X \pm \sqrt{1 - X^2}, \quad X \leq 1
\end{equation}
The  two $(X \geq 1)$, $(X \leq 1)$ domains being the classical and quantum domains 
respectively with their two $(\pm)$ branches each, and when $ 
x_{+} = x_{-} :  X = 1,\;  x_{\pm} = 1$,  (Planck scale).

\medskip

The two  domains precisely account for the two different 
and duals ways of approaching the Planck mass:
The elementary particle domain $0 \leq x \leq 1 $,
 and the macroscopic gravity domain  $1 \leq  x \leq \infty$. 
These two domains are duals of each other in the precise sense of 
the classical-quantum duality through the "Planck scale duality" Eqs (1.5), (1.6).
In other words:

\medskip

Quantum theory has $L_Q >> l_P$ \quad and \quad $L_G << l_P$    

\medskip

Classical gravity has $L_Q << l_P$ \quad and \quad $L_G >> l_P$

\medskip

Quantum Gravity has $L_{QG}$ \; and \; any value of $ L_G$ and $L_Q$, 
and the Planck domain as well.

This is reflected in the QG duality satisfied intrinsically by the QG variables  
$L_{QG}$ and $M_{QG}$ as well as in the "Planck duality" relation E.(1.5) 
for both Lengths $L_G$ and $L_Q$ separately and for the {\it dual (quantum)} Mass $M_Q$:
\be
 M_Q = \frac{m_P^2} {M}  
\ee
\begin{figure}
\includegraphics[height=15 cm,width=15 cm]{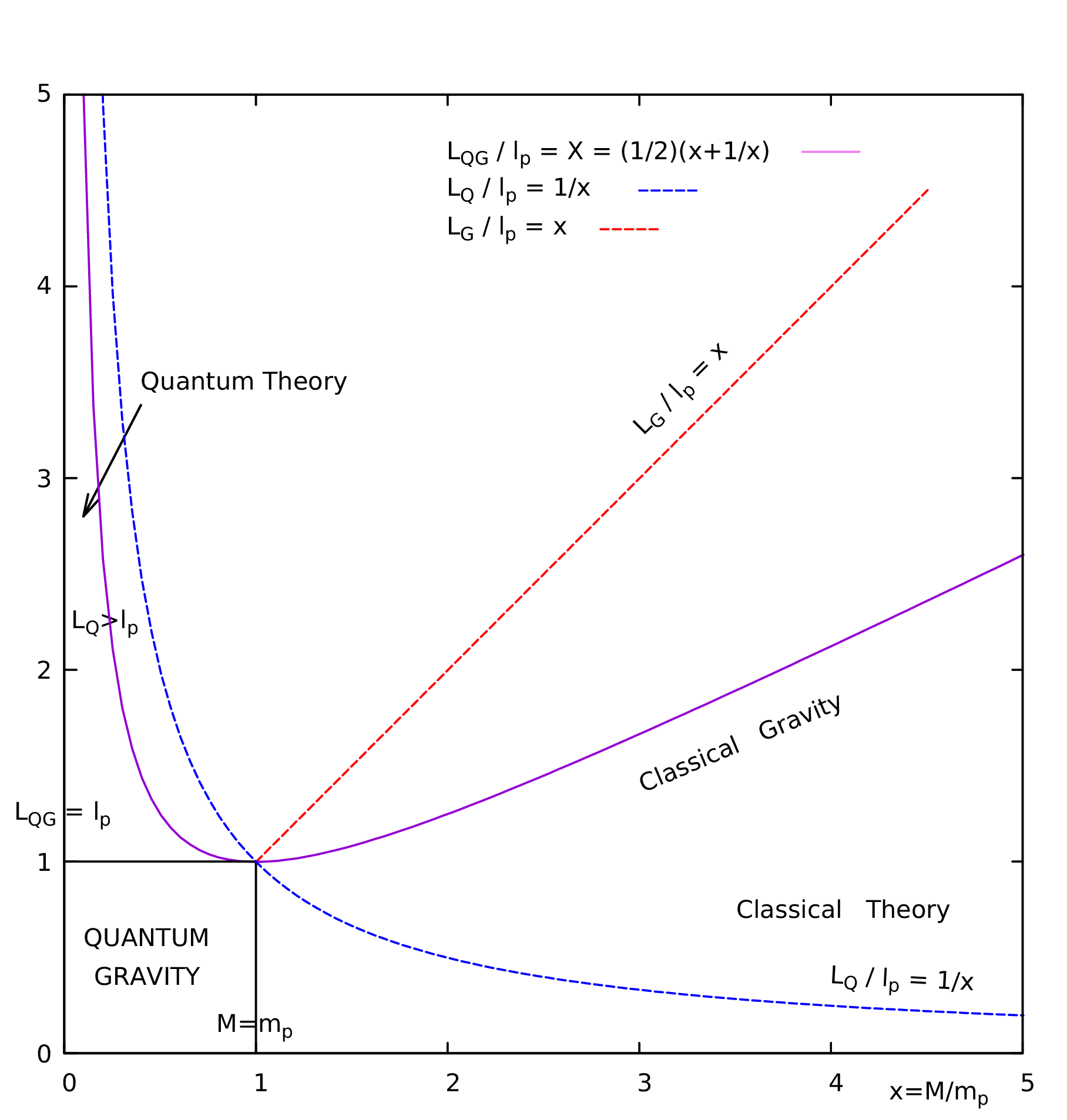}
\caption{The dimensionless QG variable $X = (1/2) (x + 1/x), X \equiv L_{QG}/l_P$ 
and its classical and quantum components against the dimensionless mass $x = M/m_p$ 
($m_P$ and $l_P$ stand for the Planck mass and Planck length). For completeness, 
each component of $L_{QG}/l_P$ : the dimensionless classical gravity length 
$L_G/ l_P = x$ and the quantum length $L_Q/l_P = 1/x$ are also displayed. 
$X = 1, x = 1$ at the Planck scale delimitates the QG Planck scale domain.}
\label{fig1}
\end{figure}
The elementary particle domain of masses and lengths Eq.(2.4)
is mapped respectively through Eq.(1.5) into the macroscopic mass and 
length domain of classical gravitational objects Eq.(2.5)
and conversely. More generically: A  set of classical gravitational quantities 
$$
O_G =(L_{G},\, M,\, K_{G},\, T_{e})
$$
 characteristic of the classical gravity regime, 
 (here denoting length, mass, gravity acceleration, temperature respectively), 
and the corresponding set in the quantum regime 
$$
O_Q=(L_Q,\, M_{Q},\, K_Q,\, T_{eQ})
$$
are {\bf dual} of each other (in the precisely sense of the  classical- quantum duality), ie
\be
O_{G}=o_{P}^2\, O_Q^{-1}
\ee
This relation holding for any quantity in the set. $O_{G}$ and 
$O_Q$ being the same conceptual physical quantities in the different 
(classical and quantum) regimes including gravity. 
$o_{Pl}$ are the corresponding Planck scale magnitudes
depending only of ($\hbar ,\, c,\, G$). $O_Q$ stands for the quantum magnitudes: 
quantum size $L_Q$, quantum mass $M_Q$, 
quantum surface acceleration $\mathcal{K}_Q$ and quantum temperature $T_{eQ}$:
$$ T_{eQ}=\frac{1}{2\pi k_B}\, M_Qc^2 $$
The energy $M_Q c^2$ has naturally associated a quantum temperature 
(in units of $ k_B$), which is also 
a measure (in units of $k_B$) of the quantum length $L_Q$, and $K_Q = c^2/L_Q$. 
In particular, 
for semiclassical black holes, $T_{eQ}$ is the Hawking temperature, 
but $T_{eQ}$ is a much more generic concept.

\medskip

There is thus a {\it duality} between the quantum domain of elementary particles 
(without gravity) and the classical gravity domain of macroscopic masses (without the quantum 
theory). 
In other words: The quantum domain of elementary particles 
(at the level of the set of fundamental variables as masses, lengths, temperatures, 
acceleration)
is {\it dual} (in the precise sense of classical-quantum duality) 
of the classical gravity domain. This is not an assumed or conjectured duality 
and does not relate to the number of 
dimensions or space-time compactification, nor to any imposed symmetry.

\section {Complete Analytic Extension of the QG Variables}

Another QG variable, a "time"- type variable $T_{QG}$ can be formed from
the  difference of $L_G$ and $L_Q$  (and similarly, a QG momentum 
 with $M_G$ and $M_Q$):

\begin{equation}  
T_{QG}= \frac{1}{2} \;(T_G - T_Q), \; \qquad T_G = \frac{L_G}{c}, 
\quad \; T_Q = \frac{L_Q}{c} = \frac{t_P^2}{T_G}           
\end{equation}
$$
T  =  \frac {1}{2} \;(t - \frac{1}{t}), \quad \;  T\equiv \frac{T_{QG}}{t_P}, 
\quad \; t \equiv \frac{T_G}{t_p},
$$

$t_{P}$ being the Planck time scale \; $t_{P}^2 = \hbar G / c^5 $. 
The QG variables $X$ and $T$ satisfy:  
$$
X(x) = X({x}^{-1}),  \quad X(-x) = -X, \qquad  X(\pm\infty) = \pm\infty,
\quad X(0) = \infty ,\quad X(1) = 1
$$
$$
T(t) = -T ({t}^{-1}) ,  \quad  T(-t) = -T ,\qquad   T(\pm\infty) = \pm\infty,
\quad T(0) = -\infty ,\quad T(1) = 0
$$
These properties of the $(X,T)$ variables are more apparent in terms of the "star- 
coordinates" $(x{\ast}$, $t{\ast})$:
$$
x = \exp{ (\kappa x\ast)},  \quad  t = \exp{(\kappa t\ast)}, \quad \mbox{which imply} 
\quad \;  X = \cosh {(\kappa x^*)}, \quad  T= \sinh {(\kappa t^*)}.
$$
The parameter $\kappa$ is a pure number introduced for convenience of the sequel, 
it plays the role of the dimensionless surface gravity (gravity acceleration)
allowed by the mapping. (In particular, at the Planck scale: $\kappa = 1$).
\medskip

This leads us to consider `null-type'  QG variables and a more general exponential mapping
yielding the QG variables $(X,T)$, in terms of $(x*, t*)$:  
\begin{equation} 
X \pm T = \exp{{{\kappa(x* \pm t*)}}}, \qquad U = X - T ,\quad V = X + T,  
\end{equation}
which maps characteristic lines into characteristic lines,
ie the null type structure of $(X, T)$ and $(x*, t*)$ is preserved, 
it is analytic and holomorphic if continuated to complex variables. 
In the $(x, t)$ variables, this mapping simply reads:
$$
U= x \; t^{-1}, \;  \quad V = x \; t
$$
The $QG$ variables $(X,T)$ are complete coordinates fully covering 
the analytic extension of the QG manifold, while $(x*,t*)$ or $(x, t)$
 cover incomplete domains. $(X,T)$ are "Kruskal"-like variables, while 
$ (x\ast, t\ast$) or $(x, t)$ are "Schwarzschild"-like  or "Rindler"-like    
since they are bounded, local or covering incomplete domains. 
Still, the comparison with the Schwarzschild $r$-coordinate is more apparent 
by singling out the role of  $x = 1$ (Planck scale) in this context:
$$
x* = r + \frac {1} {\kappa} \log \sqrt{ 2 \kappa r - 1} , 
\qquad    1 < 2 \kappa r \leq \infty, \ \quad \  -\infty \leq  x* \leq \infty
$$
$$
\mbox{Thus,}\quad x =  \sqrt{2 \kappa r - 1}\; \exp {(\kappa r)},  
\; \quad  0 \leq x \leq \infty, 
 \quad \mbox{and}  \quad   1 \leq t \leq \infty, \; \quad \;  -\infty \leq  t* \leq \infty
$$
Four charts I, II, III, IV are necessary to fully describe the complete QG domain.  
It follows:
\begin{equation}
X = \exp{(\kappa x*)} \;\cosh {(\kappa t*)} = 
\sqrt{2\kappa r - 1}\;\exp{(\kappa r)}\;\cosh {(\kappa t*)}  
\end{equation} 
\begin{equation}
T = \exp {(\kappa x*)} \;\sinh {(\kappa t*)} = 
\;\sqrt{2\kappa r - 1}\; \exp{(\kappa r)}\; \sinh {(\kappa t*)}
\end{equation}
$$
UV = X^2 - T^2 = e^{(2\kappa x*)} = (2\kappa r - 1)\; e^{(2\kappa r)}, \quad \;
\frac {V}{U} = e^{(2\kappa t*)}, \quad \; \frac {T}{X} = \tanh (\kappa t*)
$$
\begin{figure}
\includegraphics[height=14.cm,width=16.cm]{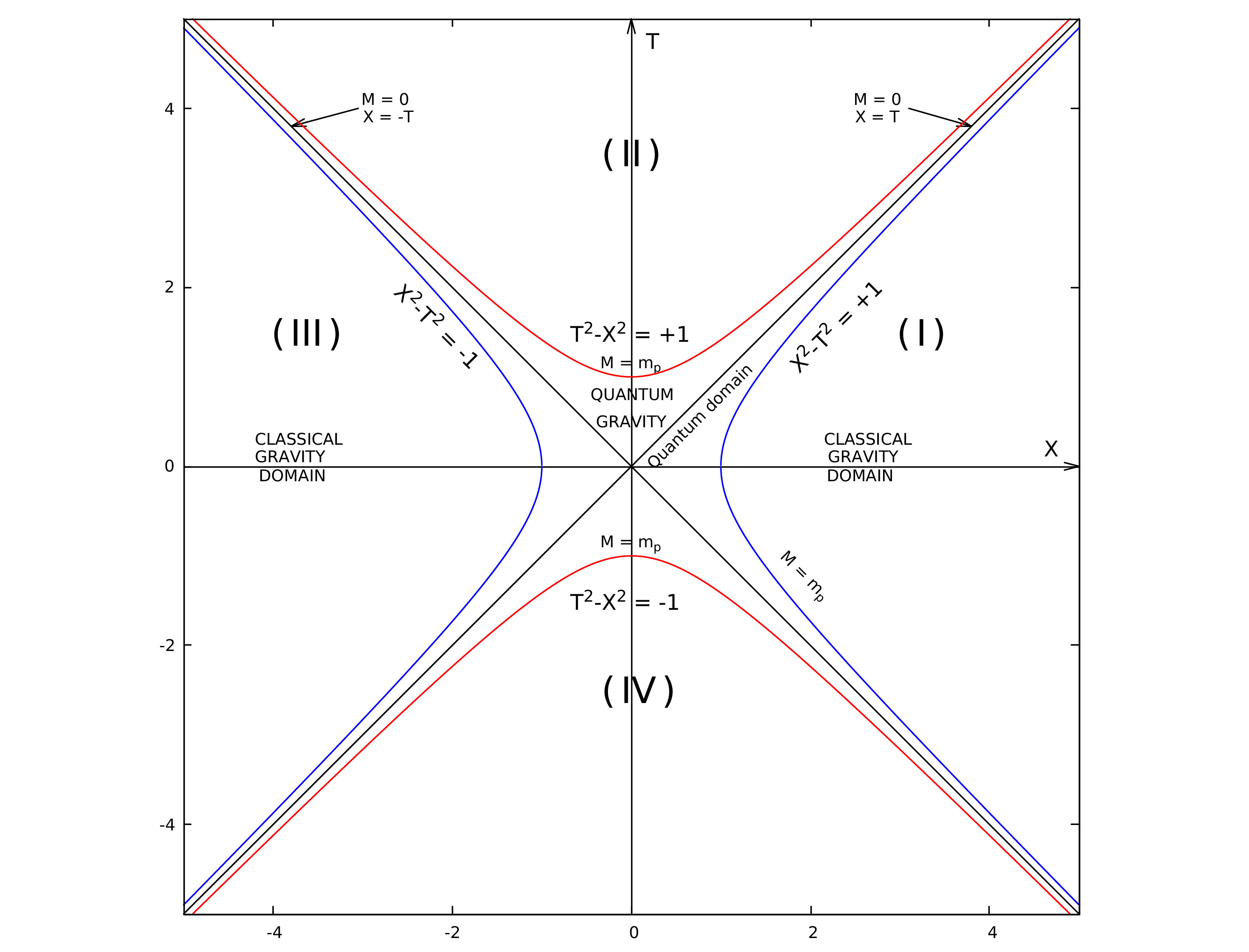}
\caption{{\bf The complete analytic extension of the QG variables 
$(X, T)$}. Four regions are needed to cover the full manifold QG. 
$(X, T)$ are complete variables covering
 all the classical and quantum domains and the quantum gravity domain 
 as well, while $(x,t)$ or $(x*, t*)$ are  incomplete variables, covering 
 only one region (classical or quantum) without covering the 
 quantum gravity domain. The four Planck scale hyperbolae $X^2 - T^2 = \pm 1, 
  \; T^2-X^2 = \pm 1$ (in Planck units) delimitate the quantum region including the quantum gravity 
 domain (quantum range  of masses -as elementary particles- passing by zero to the 
 Planck mass). $1$ is here $l_P^2$. The Planck mass is reached from the classical side 
 of large masses including the largest astronomical objects (region I and its mirror III) and from the 
quantum side of small masses (regions II and IV), as it must be: (classical-quantum duality at work through the Planck scale).}
\label{fig2}
\end{figure}
Or, in terms of the $(x,t)$ variables, these equations read:
\begin{equation}
X  =  \frac {x}{2}\; (t + \frac{1}{t}), \qquad  \qquad 
T =   \frac {x}{2}\; (t - \frac{1}{t})                      
\end{equation}
\be
UV = x^2, \qquad
\frac {V}{U} = t^2 , \qquad \frac {T}{X} = \frac {t - t^{-1}}{t + t^{-1}}
\ee

Equations (3.3)-(3.4) correspond to the domain (I):  
$\left| X \right| > T$ , $ x > 0 $, ($ 2\kappa r > 1$) and its mirror copy 
$\left| X \right| < T$, (region III). .
These  are the {\it "exterior" regions} $ 1 < x \leq \infty $, 
that is  $L_G$ and $M$  larger than $l_P$ and $m_P$ respectively, 
i.e the classical and semiclassical regions. 
They include the known classical gravity General Relativity domain, 
the semiclassical gravity and quantum theory (without quantum gravity) 
domain of the incomplete variables $(x,t)$. 
The complete $QG$ ($X,T$) variables can take all values 
from $0$ to $\pm$ $\infty$ passing by $1$ without any "boundary" or 
restriction.

\medskip

The "past" and "future" QG {\it interior regions} $T > \left| X \right|$ 
and $T < \left| X \right| $ 
are precisely the {\it full quantum} domains where $ x \leq 1$, ie $L_G \leq l_P$, 
$M \leq m_P$ (regions II and IV respectively).  
In these QG domains, $ (X, T)$ are given by the "interior" variables, 
which are the analogues of Eqs (3.3)-(3.4) with $ X$ and $T$ interchanged, 
and $ x*$ (or $x$) given by:
$$
x* = r + \frac{1} {\kappa}\log\sqrt{1-2 \kappa r}, \quad \;
x = \sqrt{1-2 \kappa r}\; \exp (\kappa r), \quad \; 0 \leq 2\kappa r \leq 1, 
\; \quad  0 \leq x \leq 1 
$$
\begin{equation}
X = \exp{(\kappa x*)}\; \sinh {(\kappa t*)} = 
\frac {1}{2} \; \sqrt{ 1- 2\kappa r }\;e^{(\kappa r)} \; (t - \frac{1}{t}) 
\end{equation}
\begin{equation}
T = \exp{(\kappa x*)} \;\cosh {(\kappa t*)} = 
\frac {1}{2} \; \sqrt{ 1- 2\kappa r }\;e^{(\kappa r)} \; (t + \frac{1}{t}) 
\end{equation}
\be
UV = T^2 - X^2 = e^{(2\kappa x*)}  =  x^2 ,\qquad
V/U =  e^{(2\kappa t*)} = t^2
\ee

The fixed mass values $x = $const.  follow the trajectories $X^2 - T^2$ = const.
The straight lines $X = \pm T$ , ($U = 0$ or $V = 0$) for
 $x* = -\infty$ , (that is $ 2\kappa r = 1$) are at the null values $x = 0: 
 L_G = 0, M = 0$, (null (future or past) horizons:  $t * = \pm \infty$, (being $t = + \infty$ or $0$). 

\medskip

The Planck scale $x = 1$ (ie $M = m_P$) corresponds to the hyperbolae 
$T^2 - X^2 = \pm 1$ at $r = 0$ (ie $x* = 0$):
\; $x = 1 $ in the future (+), and past (-) interior regions. 
The correspondence with the Kruskal manifold is manifest.

\section{Classical-Quantum Duality of the Schwarzschild-Kruskal Space-Time}

Let us now consider the Schwarzschild black hole: 
$$
dS^2 = -(1 - 2 \kappa /r) \;dt*^2  + (1 - 2 \kappa /r)^{-1} dr^2 + d\Omega_{\bot}^2 
$$
Here $(r, t*)$ are the usual Schwarschild space-time coordinates,  
$x* = r + \frac{1} {\kappa}\log\sqrt{1- 2\kappa r} $ and  
$\kappa =  c^4 /(2 r_{BH}) =  c^4/ (4 G M_{BH} )$, 
$M _{BH}$ being the black hole mass. 

The black hole Kruskal mapping is the same 
as Eq (3.2) and the space-time Kruskal coordinates
have identical expressions to Eqs (3.3)-(3.4): 
\be
X \pm T = \exp{\kappa(x* \pm \;t*)}, \qquad  U = X - T,\;\quad V = X + T
\ee
In the interior black hole regions  $0 \leq 2\kappa r \leq 1$:   
$x* = r + \kappa^{-1}\log\sqrt{1 - 2\kappa r} $,
$ (X,T)$ are interchanged and are given by Eqs (3.7)-(3.9). 
 
\medskip

In terms of the $(x,t)$ coordinates, Kruskal $(U,V)$ coordinates 
read simply
\be
U =  {x}\; t^{-1}, \qquad \quad
V =  {x} \; t, \qquad
UV =  {x}^2,   \qquad \frac {V}{U} = t^2 
\ee
As is known, $ X^2 - T^2 = 0 $ are the past and future null horizons 
$2\kappa r = 1$, $(x*=\infty)$, here at ${x} = 0$.
$T^2 - X^2 = \pm 1$ are the past $(-)$ and future $(+)$ 
classical space-time singularities at $ r= 0 $,  $(x*= 0)$, here at ${x}= 1$.

\begin{figure}
\includegraphics[height=14.cm,width=16.cm]{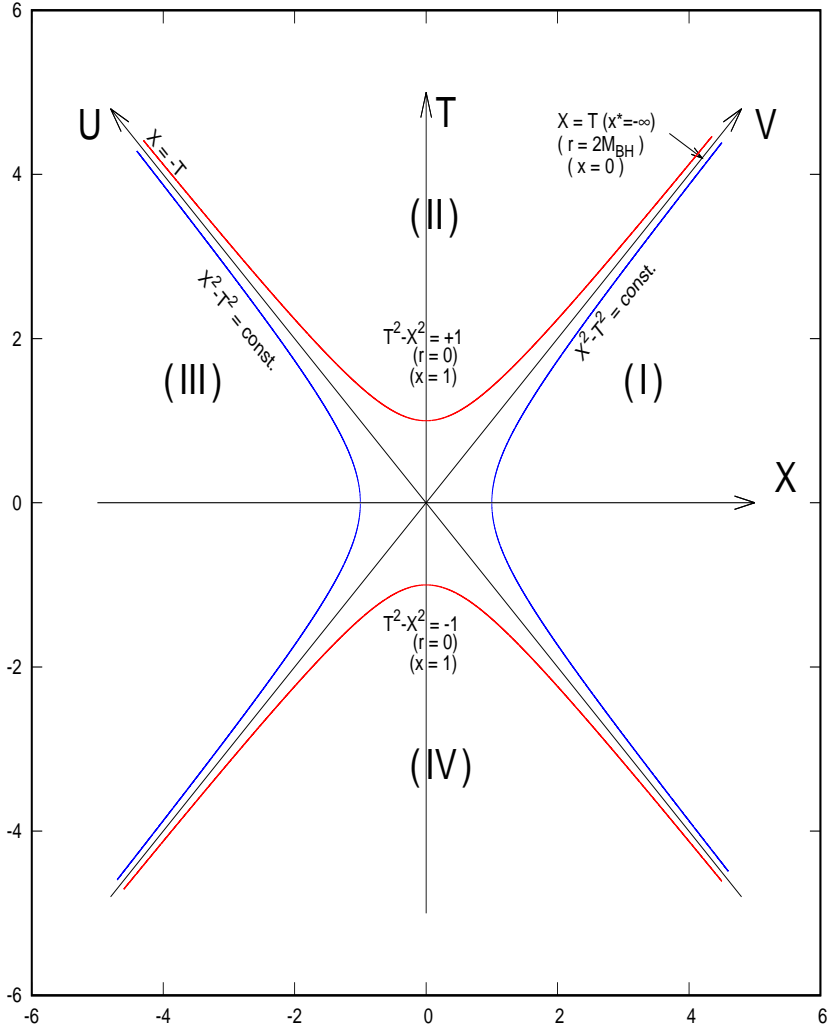}
\caption{{\bf The complete Schwarzschild-Kruskal 
space-time and its classical-quantum duality}. Kruskal $(X,T)$ 
coordinates are QG 
variables. Exterior  (I) and (III) regions are classical or semiclassical. The interior
regions (II) and (IV) are fully {\it quantum}.  The four Planck scale hyperbolae $X^2 - 
T^2 = \pm 1, \; T^2-X^2 = \pm 1$ (in Planck units) delimitate the quantum gravity region 
{\it including} the horizons $X = \pm T$. 
Exterior and interior lose their difference near the horizon:
The horizons are bordered by the four Planck scale hyperbolae which act like a {\it quantum dressing} or 
width, we call it {\it l'horizon habill\'e} (dressed horizon) . $1$ is here $l_P^2$. 
The Planck scale is reached from the exterior: $ X^2 -T^2 = \pm 1$ and from the 
interior: $T^2-X^2 = \pm 1$ \;(at $r=0$). The antipodal
space-time symmetry and the PT or CPT symmetry are contained in the QG symmetry.
Other properties and symmetries of the 
manifold as classical-quantum duality mappings are discussed in the text.}  
\label{fig3}
\end{figure}

\medskip

The Kruskal-Schwarzschild manifold is 
endowed with several discrete symmetries
\be
U \rightarrow \pm V ,\quad V \rightarrow \pm U, \quad \mbox{ie} 
\quad X = - X,\quad T = - T, 
\ee
which can be accompagnied by corresponding changes in the transverse 
(angular) coordinates and their antipodal ones 
\be \Omega_{\bot}\rightarrow \pm \Omega_{\bot}.\ee  
The {\it duality symmetry} of the Kruskal space-time (hidden by 
the usual Schwazrschild coordinates $(r, t*)$)  
shows up clearly in the $(x,t)$ variables:
\be
X (t^{-1}) = X (t),\qquad  T (t^{-1}) = - T (t)  
\ee
\be
U (t^{-1}) = V (t),\qquad   V (t^{-1}) = U (t)
\ee
This duality under $ t \rightarrow t^ {-1}$ means that the time domains:
$0 \leq t \leq 1 $ and $ 1 \leq t \leq \infty$ are mapped into
$\infty \leq t \leq 1$ and $1 \leq t \leq 0$ respectively,
or equivalently, into  $-\infty \leq t* \leq 0$ and $ 0 \leq t* \leq \infty$ .
Recall here that $t$ is dimensionless (in units of the Planck time $t_P$) and arises 
from a physical classical gravitational time $ T_G$ as in Eq.(3.1), or a classical length 
$ L_G$ which is in turn connected to the quantum length $L_Q$ and thus 
to the quantum dual time $T_Q$ as given by Eqs(3.1). 
Thus, in this context, the duality transform  $t\rightarrow{t}^{-1}$ changes the 
classical time $t$ into the quantum time $t^{-1}$ and conversely,
but the Kruskal time $T$ as the quantum gravity (QG) time remains invariant. 
The  external regions (I) and (III) are classical gravity and semiclassical  
domains while the interior regions are quantum. These regions are {\it dual} 
of each other in the precise sense of the classical-quantum duality.

\bigskip

The classical-quantum duality transform in the Kruskal space-time maps the internal 
quantum region  $x \in (0,1)$ into the external classical domain 
$x \in (1, \infty)$ through the Planck scale $(x = 1 , t = 1)$ and conversely. 
The halves (I, II) and (III, IV), 
ie the classical/semiclassical wedge (I) and its corresponding quantum (future) interior II
on one side, and the copy III and its corresponding (past) region IV on the other side, 
are antipodally identical of each other.
This supports the {\it antipodal identification} of the two worlds ref [7], 
refs [8],[9].
By promoting the QG variables to quantum (non conmutative)
coordinates, further insight into the 
quantum space-time structure is obtained and the 
identification of the two Kruskal copies is further supported by the quantum theory 
ref [10].

\bigskip

In the interior regions the Planck scale is reached at the loci $T^2 -X^2 = \pm 1$ 
($r=0)$ which are bordering the null horizons $X = \pm T$ (at $2 \kappa r = 1$) 
and asymptotically reaching them for $T\rightarrow  \pm \infty $, Fig 3.
The {\it quantum region extends beyond the horizons} and is delimitated by 
the Planck scale hyperbolae $X^2 -T^2 = \pm 1$. 
These hyperbolae are acting like a {\it quantum dressing} or quantum 
width for the horizon, let us call it {\it l'horizon habill\'{e}} (dressed horizon). 
$1$ is here $l_P^2$.  We see that exterior and interior lose its difference
near the horizons $X =\pm T$. 
Although, is possible to use these words in scales larger than the Planck scale.

There are {\it two} dual ways of reaching the horizon and the Planck scale: from the exterior 
(classical/semiclassical) and from the interior (quantum) regions. This appears like a 
{\it"splitting" or shifting} of the null horizons $X = \pm T$ 
into the hyperbolae $X^2 -T^2 = \pm 1 , X = \pm \sqrt{T^2 \pm 1}$ being 
 $1$ the (dimensionless) Planck scale.   
We find a similar result when promoting $(X,T)$ to quantum non-commutative coordinates 
as we do in ref [10].
 
 \medskip

Let us recall here
that quantum field or string dynamics near horizons
require quantum "shifts" or widths of the horizon 
of the order of the Planck scale, refs [11], [12],[13]. 
For quantum fields, the quantum shift appears as produced by the back reaction and the dynamics 
of ingoing and outgoing modes crossing the horizon ref [11]. 
For quantum strings, the horizon width is required by the size (of the order of the Planck scale) 
of the string . We find interesting that in 
a different approach, and {\it solely} from our QG variables, a horizon {\it quantum dressing} or width 
does appear, (and without searching for it) supporting the quantum nature of this space-time region.
 The QG variables naturally contain the Planck scale and in some sense the
horizon splitting due to the {\it two} valued and dual covering (like Kruskal
coordinates). Other related results are reported in ref [10].

\section{Concluding Remarks}

\begin{itemize}
\item{We have provided here new quantum gravity (QG)
variables which extend the known gravity (G) and quantum (Q) variables
and reduce to them in each of the limiting sectors $G >> Q$ or $Q >> G$.

\medskip

The QG variables contain together both classical and quantum domains
including quantum gravity and the elementary particle sector passing 
by the Planck scale. 
They generalize the gravity (G) and quantum (Q) variables of each 
classical and quantum domain, and contain their duality properties. 
They are invariant under the quantum (Q) and gravity (G) duality 
through the Planck scale: $Q\rightarrow G^{-1}$ duality, and conversely
$G \rightarrow  Q^{-1}$.}
 
 \item {The dual of classical behaviour is quantum behaviour:   
"de Broglie duality'' with $\hbar $, "Compton duality'' with $\hbar $ and $c$, 
 our " Planck duality'' with $\hbar, c$ and $G$: 
$O_Q= o_P^2 O_G^{-1}$,  
and "QG duality" here  which extends all of them: \\
$O_{QG}\; (O_G) = O_{QG}\; (O_Q)$, ie $O_G \rightarrow O_Q$ and conversely.

\medskip

The QG variables $O_{QG} = (1/2) (O_G + O_Q )$ naturally contain 
all the above dualities.
This duality is {\it universal}. As the wave-particle duality,
it does not relate to the number 
or the kind of dimensions nor to any imposed symmetry or compactification. 
QG variables are automatically dual symmetric, $O_G \rightarrow O_Q$:\\
$O_{QG}\;(1/x) = O_{QG}\;(x), \; x = O_G/o_P = o_P/O_Q$.}

\item{The complete analytic extension of QG variables allowed us to reveal 
the classical- quantum duality of the space-time.  

The regions I and III (Fig 2): exterior to the Planck scale hyperbolae 
$X^2-T^2 = \pm 1$
contain both classical and semiclassical behaviours, 
depending on whether the classical or the quantum component or regime dominates, 
that is  whether $O_G$, or $O_Q$ are dominant, while the interior regions II and IV 
(inside the four Planck scale hyperbolae) are totally {\it quantum} 
in the range  $[0, l_P]$.}

\item {This sheded light on new quantum properties of the 
Schwarschild-Kruskal black hole structure.

\medskip

The four Planck scale hyperbolae $X^2 - T^2 = \pm 1$, $T^2-X^2 = \pm 1$ 
(in Planck units) delimitate the quantum gravity region.  
The horizons $X = \pm T$ are now
bordered by the four Planck scale hyperbolae which are like 
a {\it quantum dressing} or quantum width
for the horizon,  {\it l'horizon habill\'{e}} ("dressed horizon"). 
$1$ is here $l_P^2$. 

The Planck scale is reached from the exterior $X^2-T^2 = \pm 1$ 
and from the interior: $r=0$ is the Planck scale $T^2-X^2 = \pm 1$.

\medskip

The exterior and interior regions thus appear {\it "dressed" and redefined}: 
they acquire a Planck scale structure. 
Near the horizons $ X = \pm T$,  exterior and interior 
{\it lose their difference}.

\medskip

We have already discussed in the Introduction and along the paper 
the main new features of the paper and will not include all of them again here.
We refer to section I for a summary of the results.}

\item {The antipodal identification of space-time ("aist" in short) 
is not the purpose nor the subject of this paper, but as already mentionned, 
our results here support aist in the quantum theory, and 
our new quantum space-time structure results ref [10] imply it.
We include then here some remarks to update the issue given the new context here and the recent
 refs [8],[9]. In refs [7],
we investigated aist in a semiclassical QFT description, imposing such a boundary condition to 
the vacuum in a "strong way" stressing that it would give a zero norm state in a global spatial 
section. But in each halve it is perfectly consistent to construct antipodally symmetric (or 
antipodally antisymmetric) states with non-zero norm. (Also, the classical global Cauchy section 
would not be such that in full quantum gravity).
Importantly, in aist the space-time topology changes and the resulting manifold (coset or 
quotient space) is projective,
another characteristic manifestation (non trivial topology) of quantum gravity. In aist 
the normalized antipodally symmetric (or antisymmetric) theory has {\it two} times more 
probability than
the non aist symmetric theory. These differences of factors two in the aist and non-aist 
theories are {\it not} 
misleading, they are totally correct and perfectly understandable in the two-fold covering of 
the global manifolds. The QG variables and QG symmetry encloses all that.}

\item{The size of the black hole is the gravitational length $ L_G $ in the 
classical regime, it is the Compton length $L_Q$ in the semiclassical regime, it is the 
Planck size (or the string size $L_s$) in the full quantum gravity regime. Similarly,
the horizon acceleration (surface gravity)  of the black hole is 
the Planck  acceleration $\kappa_P$ in the QG regime. The  
Hawking temperature $T_{eQ}$  (measure of the surface gravity or of the 
Compton length)  becomes the Planck temperature  
in the full quantum gravity regime. The gravitational thermal features as Hawking radiation
are typical of the semiclassical phase. The end of evaporation is non thermal
and purely quantum. For masses smaller than the Planck mass the black hole {is not
\it anymore} a black hole but a elementary particle state.
Moreover, the quantum mass spectrum we found recently for {\it all} masses
confirms this picture ref [10].}

\item {Space-time can be parametrized
by {\it masses} ("mass coordinates"), just related to length and time as  QG variables, 
on the same footing of space and time.
In Planck units, any of these variables (or another convenient set) can be used. This
reveals particularly interesting for mass quantization ref [10].

\medskip

More generally, other  analytic mappings for full analytic extensions 
in other manifolds could be considered, ref[14].
In QG, functions of the QG variables 
could satisfy QG duality symmetry: 
$ f(x,t) = \pm $ const. $ f (x^{-1}, t^{-1})$ . 
QG variables can be also considered in phase-space 
with their full analytic extension to all values or patches.} 

\item {In summary, {\it QG variables} or coordinates cover 
all the domains, classical and quantum, with and without gravity. 
They turn out to be the classical-quantum duals of each other, 
in the precise sense of the wave-particle duality, here extended 
to the quantum gravity (Planck) domain: 
wave-particle-gravity duality: {\it QG duality} in short. 
QG variables can be used directly as space-time variables 
or coordinates as we do here, 
they can be considered as the result of expectation values in a quantum state 
or they can be used to construct appropriate quantum operators 
as we do in ref [10].}

\item{The idea of a continuum space-time is a classical (non-quantum) 
approximation. A step forward in which the space-time itself is quantized 
confirms the results presented here and yields more radical new results 
presented in another paper ref [10].}

\end{itemize}

\bigskip

{\bf ACKNOWLEDGEMENTS}

\bigskip

The author thanks G.'t Hooft for interesting and stimulating communications 
on several occasions,
M. Ramon Medrano for interesting discussions and encouragement and F. Sevre for help with the 
figures. The author acknowledges the French National Center of Scientific Research (CNRS)
for Emeritus Director of Research contract. This work was performed in LERMA-CNRS-Observatoire 
de Paris-PSL Research University-Sorbonne Universit\'{e} UPMC.

\end{document}